\documentclass[aps,prd,twocolumn,preprintnumbers,superscriptaddress,showpacs,floatfix,nofootinbib]{revtex4}

\usepackage{bm}
\usepackage{epsfig}
\usepackage{graphics}
\usepackage{graphicx}

\newcommand{\lsim}{\mbox{\raisebox{-.9ex}{~$\stackrel{\mbox{$<$}}{\sim}$~}}}
\newcommand{\gsim}{\mbox{\raisebox{-.9ex}{~$\stackrel{\mbox{$>$}}{\sim}$~}}}

\renewcommand\({\left(}
\renewcommand\){\right)}

\newcommand\eq[1]{Eq.~(\ref{#1})}
\newcommand\eqs[2]{Eqs.~(\ref{#1}) and (\ref{#2})}
\newcommand\eqss[3]{Eqs.~(\ref{#1}), (\ref{#2}), and (\ref{#3})}

\newcommand\eqref[1]{(\ref{#1})}

\newcommand\ee{\end{equation}}
\newcommand\be{\begin{equation}}
\newcommand\eea{\end{eqnarray}}
\newcommand\bea{\begin{eqnarray}}

\newcommand\mpl{M_{\rm P}}

\def\calh{{\cal H}}

\def\calp{{\cal P}}

\newcommand\bfk{{\bf k}}

\newcommand\bfx{{\bf x}}

\newcommand\Mpc{\,\mbox{Mpc}}

\newcommand\msun{M_\odot}

\newcommand\sub[1]{_{\rm #1}}
\newcommand\su[1]{^{\rm #1}}

\newcommand{\one}{_1}
\newcommand{\two}{_2}

\newcommand\mone{^{-1}}
\newcommand\mtwo{^{-2}}

\newcommand{\vev}[1]{{\langle #1 \rangle }}

\newcommand{\fnl}{f\sub{NL}}

\begin{document}

\title{Non-gaussianity from the second-order cosmological perturbation}

\author{David H. Lyth}
\email{d.lyth@lancaster.ac.uk}
\affiliation{Department of Physics, Lancaster University, Lancaster LA1 4YB, UK}
\author{Yeinzon Rodr\'{\i}guez}
\email{y.rodriguezgarcia@lancaster.ac.uk}
\affiliation{Department of Physics, Lancaster University, Lancaster LA1 4YB, UK}
\affiliation{Centro de Investigaciones, Universidad Antonio Nari\~no, Cll 58A \# 37-94, Bogot\'a D.C., Colombia}
 
\date{\today}

\begin{abstract}
Several conserved and/or gauge invariant quantities described as the second-order
curvature perturbation have been given in the literature. We revisit various scenarios
for the generation of second-order non-gaussianity  in the primordial 
curvature perturbation $\zeta$, employing for the first time a unified notation and
focusing on the normalisation $f\sub{NL}$ of the bispectrum. When $\zeta$
 first appears a few Hubble times after horizon exit, $|f\sub{NL}|$
is much less than $1$ and is, therefore, negligible. Thereafter $\zeta$ (and hence $f\sub{NL}$)
is conserved as long as the pressure is
a unique function of energy density (adiabatic pressure).
Non-adiabatic pressure comes presumably only from the effect of 
fields,  other than 
the one pointing along the inflationary trajectory, which are light during inflation
(`light non-inflaton fields').
During single-component inflation $\fnl$ is constant, 
but multi-component inflation might generate $|f\sub{NL}|\sim 1$ or bigger.
Preheating  can  affect $\fnl$ only in atypical scenarios where 
it involves light non-inflaton fields. The simplest  curvaton 
scenario typically gives $\fnl\ll -1$ or $\fnl=+5/4$. 
The inhomogeneous reheating scenario can give a wide range of values for $\fnl$.
Unless there is a detection, observation 
can eventually provide a limit $|\fnl|\lsim 1$, at which level it will be crucial to 
calculate  the precise observational limit using second order theory.
\end{abstract}

\maketitle

\section{Introduction}

Cosmological scales
leave the horizon during inflation and re-enter it after Big Bang 
Nucleosynthesis. Throughout the super-horizon era
 it is very useful
to define a primordial cosmological curvature perturbation,
 which is 
conserved if and only if  pressure throughout the Universe is 
a unique function of energy density 
(the adiabatic pressure condition) \cite{bardeen,bst,lyth84,sb,wands,rs1,sasaki1}. 
Observation directly constrains the  curvature perturbation
at the very end of the super-horizon era,
a few Hubble times before cosmological scales start to enter the horizon,
when it apparently  sets the initial condition for the subsequent evolution of
all cosmological perturbations. The observed curvature perturbation
is almost Gaussian with an almost
scale-invariant spectrum. 

Cosmological perturbation theory expands the exact equations in
 powers of the perturbations
and keeps terms only up to the $n$th order. Since the observed
curvature perturbation 
is  of order $10^{-5}$,
one might think that first-order
perturbation theory will be  adequate for all comparisons with observation. 
That may not be the case however, because 
the PLANCK satellite \cite{planck} and its successors may  be sensitive to non-gaussianity
of the curvature perturbation at the level of second-order perturbation
theory \cite{spergel}. 

Several authors have treated the non-gaussianity of the primordial curvature
perturbation in the context of
second-order perturbation theory. They have adopted different definitions 
of the curvature perturbation and 
obtained  results for a variety of situations.
In this paper we  revisit  the calculations, using a single
definition of the curvature perturbation
 which we denote by $\zeta$. In some cases we disagree with the findings of 
the original authors.

The outline of this paper is the following:  in Section \ref{definitions} 
we review two definitions of the curvature perturbation found in the literature,
which are valid during and after inflation,
and establish definite relationships between them; in section \ref{thirddef} we review
a third curvature perturbation definition, which applies {\it only} during inflation,
and study it in models of inflation of the slow-roll variety;
in Section \ref{origin} we describe the present framework for thinking about
the origin and evolution of the curvature perturbation;
in Section \ref{gaussianity} we see how non-gaussianity is defined
and constrained by observation;
in Section \ref{initial} we
study the initial non-gaussianity of the curvature perturbation, 
a few Hubble times after horizon exit; in Section \ref{after}
we study its  subsequent  evolution according to some different models.
 The conclusions
are summarised  in Section \ref{conclusions}. 

We shall denote unperturbed quantities by a subscript $0$,
and generally work with conformal time $\eta$ defined by the unperturbed
line element
\be
ds^2 = a^2(\eta) \( -d\eta^2 + \delta_{ij} dx^i dx^j \) \label{unpertline}
\,.
\ee
Here $a$ is the scale factor whose present value is taken to be
$1$, and  a prime  denotes   $d/d\eta$.
Sometimes though we revert to physical time $t$, 
with a dot meaning $d/dt$ defined by $d/dt\equiv a^{-1}d/d\eta$.
We  shall 
adopt the convention that a generic perturbation $g$
is split into a first- and second-order part according to the formula
\be
g =g\one + \frac12 g\two
\label{gsplit}
\,.
\ee

\section{Two definitions of the  curvature perturbation}
\label{definitions}

\subsection{Preliminaries}

Cosmological perturbations describe small departures
of the  actual Universe, away from some perfect homogeneous and isotropic
universe with the line element \eq{unpertline}.
For a  generic perturbation  it is 
convenient to make the Fourier expansion
\be
g(\bfx,\eta) = \frac1{(2\pi)^{3/2}} \int d^3k \ g(\bfk,\eta) e^{i\bfk\cdot\bfx}
\,,
\ee
where the spacetime coordinates are those of the unperturbed Universe.
The inverse of the comoving wavenumber, $k^{-1}$, 
is often referred to as the scale.



A given scale is  said to be outside the horizon during the era
$aH\gg k$, where $H\equiv \dot a/a$ is the Hubble parameter. 
{\em Except where 
otherwise stated, our discussion applies only to this super-horizon regime.}

When evaluating an observable quantity only a limited range of
scales  will be involved. The largest scale,
relevant for the low multipoles of the
Cosmic Microwave Background anisotropy,
is $k^{-1}\sim H_0^{-1}$ where $H_0$ is the present Hubble
parameter. The smallest  scale usually considered
is the one enclosing  matter with mass $\sim 10^6\msun$,
which corresponds to $k\mone\sim 10\mtwo\Mpc \sim 10^{-6}H_0\mone$.
The cosmological range of scales therefore extends over only six  
decades or so. 

To define cosmological perturbations in general, 
one has to introduce in the perturbed Universe a coordinate
system $(t,x^i)$,  which defines a slicing of spacetime (fixed $t$) and
a threading (fixed $x^i$). To define the curvature perturbation 
it is enough to define the slicing \cite{sasaki1}. 

\subsection{Two definitions of the curvature perturbation}
 
In this paper we 
take as our definition of $\zeta$ the following expression for the 
spatial metric \cite{sb,sasaki1,maldacena,zc,seery,rst,calcagni} 
which applies non-perturbatively: 
\be
g_{ij}  =  a^2(\eta) \tilde\gamma_{ij}  e^{2\zeta}
\,.
\label{mal}
\ee
Here $\tilde\gamma_{ij}$ has unit determinant, and the time-slicing
is one of uniform energy density\footnote
{It is proved in Ref. \cite{sasaki1} that this  
definition of $\zeta$ coincides with that of Lyth and Wands \cite{lyth},
provided that their slices of uniform coordinate expansion are taken
to correspond to those on which the line element has the form \eq{mal}
{\em without the factor $e^{2\zeta}$} (this makes the slices practically
flat
if $\tilde\gamma_{ij} \simeq \delta_{ij}$).}. 

It has been shown under weak assumptions \cite{sasaki1} that this
defines $\zeta$ uniquely, and that $\zeta$ is conserved as long as the 
pressure is a unique function of energy density. 
Also, it has been shown that the uniform density slicing
practically  coincides with the
comoving slicing (orthogonal to the flow of energy),   
and with the uniform Hubble 
slicing (corresponding to  uniform proper expansion, that expansion being
practically  independent of the threading which defines it) \cite{sasaki1}. 
The coincidence of these
slicings  is important since all three have been invoked
by different authors.

Since the matrix $\tilde\gamma$ has unit determinant it can be written
$\tilde \gamma = I e^h$,
where $I$ is the unit matrix and $h$ is traceless \cite{sasaki1}. 
Assuming that the initial condition is set by inflation, $h$ corresponds
to a tensor perturbation (gravitational wave amplitude) which 
will  be negligible unless the scale of inflation is very high. 
As we shall see later (see footnote \ref{teninc}), the  results we are going to present
 are valid even if $h$ is not negligible,
but to simplify the presentation we drop $h$ from the equations.
Accordingly, the space part of the metric in the super-horizon
regime is supposed to be well approximated by
\be
g_{ij} = a^2(\eta)\delta_{ij} e^{2\zeta} 
\label{expdef}
\,.
\ee

At first order, \eq{expdef} corresponds to 
\begin{equation}
g_{ij} = a^2(\eta) \delta_{ij} (1+2\zeta)
\label{all}
\,.
\end{equation}
Up to a sign, this is the definition of the first-order
curvature perturbation adopted by all authors. 
There is no universally
agreed  convention for the sign of $\zeta$.
Ours coincides with the convention of most of the papers to which we refer, and
we have checked carefully that the signs in our own set of equations are 
correct.

At  second order we have
\be
g_{ij} = a^2(\eta) \delta_{ij} (1+2\zeta+2\zeta^2)
\,.
\ee
This is our definition of $\zeta$ at second order.

Malik and Wands \cite{malik} instead  defined 
$\zeta$ by \eq{all} even at second order.
Denoting their definition by a subscript MW,
\be
\zeta\su{MW} = \zeta + \zeta^2
\label{mwdef}
\,,
\ee
or equivalently
\be
\zeta_2\su{MW} = \zeta\two + 2\( \zeta\one \)^2 \label{malmw}
\,,
\ee
where $\zeta\one$ is the first-order quantity whose definition
\eq{all} is agreed by all authors.


To make contact with  calculations of the curvature 
perturbation during inflation, we need some 
gauge-invariant expressions for the curvature
perturbation. `Gauge-invariant' means that the
definition is valid for any choice of the coordinate
system which defines the slicing and threading\footnote
{In the unperturbed limit the slicing has to be the one on which all
quantities are uniform and the the threading has to be orthogonal to it.}.

We shall write gauge-invariant expressions in terms of $\zeta$ and $\zeta\su{MW}$.
First
we consider a quantity $\psi^{\rm MW}$, defined
even at second order by
\begin{equation}
g_{ij} = a^2(\eta) \delta_{ij} (1-2\psi^{\rm MW})
\label{mw}
\,.
\end{equation}
This definition, which is written in analogy to \eq{all}, applies to a generic slicing.
Analogously to \eq{expdef} we can consider a quantity $\psi$, valid also in
a generic slicing, defined by
\be
g_{ij} = a^2(\eta) \delta_{ij} e^{-2\psi}
\label{ourpsi}
\,.
\ee 
On uniform-density slices, $\psi\one= \psi_1^{\rm MW}= -\zeta\one$,
$\psi_2^{\rm MW}=-\zeta_2\su{MW}$, and $\psi_2 = -\zeta_2$. We shall also need the 
energy density perturbation $\delta\rho$, defined on the generic slicing,
as well as the unperturbed energy density $\rho_0$.

At first order, the  gauge-invariant expression for $\zeta$
has the well-known form
\be
\zeta\one = -\psi\one - \calh \frac{\delta\rho_1}{\rho_0'}
\label{psigi}
\,,
\ee
where $\calh=a'/a$, and the unperturbed energy density satisfies
$\rho_0' = -3\calh (\rho_0 + P_0)$ with $P_0$ being the unperturbed pressure.
This expression 
obviously is correct for the uniform density slicing, and it is 
correct for all slicings because the changes in the first and second terms
induced by a change in the slicing cancel \cite{bardeen,bst,lyth84,gauget1,gauget2}.

At second order, Malik and Wands show that \cite{malik}
\bea
\zeta_2\su{MW} &=& -\psi_2^{\rm MW} - \calh \frac{\delta\rho_2}{\rho_0'}
+2\calh \frac{\delta\rho\one}{\rho_0'} \frac{\delta\rho\one'}{\rho_0'}
\nonumber \\
&&+ 2\frac{\delta\rho\one}{\rho_0'} \( \psi\one' + 2\calh \psi\one \)
\nonumber \\
&& - \( \calh \frac{\delta\rho_1}{\rho_0'} \)^2
\( \frac{\rho_0''}{\calh \rho_0'} - \frac{\calh'}{\calh^2} - 2 \)
\label{mwgi0}
\,,
\eea
which is, again and for the same reason as before, obviously correct for
all the slices.
Accordingly, from \eq{malmw}, we can write a gauge invariant definition for
our second-order $\zeta$: \footnote{This relation has recently been confirmed
in Ref. \cite{lanver} using a nonlinear coordinate-free approach.}
\bea
\zeta_2 &=& -\psi_2 - \calh \frac{\delta\rho_2}{\rho_0'}
+2\calh \frac{\delta\rho\one}{\rho_0'} \frac{\delta\rho\one'}{\rho_0'}
+ 2\frac{\delta\rho\one}{\rho_0'} \psi\one' \nonumber \\
&&- \( \calh \frac{\delta\rho_1}{\rho_0'} \)^2
\( \frac{\rho_0''}{\calh \rho_0'} - \frac{\calh'}{\calh^2} \)
\label{zetatwo}
\,,
\eea
where the relation
\be
\psi_2^{\rm MW} = \psi_2 - 2(\psi_1)^2 \,,
\ee
coming from Eqs. (\ref{mw}) and (\ref{ourpsi}), has been used.

\section{Slow-roll inflation and a third definition} \label{thirddef}

Now we specialize to the era of slow-roll inflation.
 We consider
single-component inflation, during which the curvature perturbation 
$\zeta$ is
conserved, and multi-component inflation during which it varies.
After defining  both paradigms, we give a third definition of the 
curvature perturbation which applies only during inflation.

\subsection{Single-component inflation}

In a single-component inflation model \cite{treview,book}
the inflaton trajectory is by
definition essentially
unique. 
The inflaton field $\varphi$
parameterises the distance along the inflaton trajectory.
 In terms of the field variation, slow-roll inflation is 
characterised by the slow-roll
conditions $\epsilon\ll 1$ and $|\eta_V|\ll1$ \cite{treview,book},
where
\bea
\epsilon &\equiv&  -\dot H/H^2, \label{epsilon} \\
\eta_V - \epsilon &\equiv& - \frac{\ddot\varphi}{H\dot\varphi}
\,. \label{eta}
\eea
The inflaton field 
can be taken to be canonically normalised, in which case 
these definitions
are equivalent  to 
conditions on the potential $V$
\bea
\epsilon &\equiv& \frac{\mpl^2}{2V^2} 
\( \frac{\partial V}{\partial\varphi}\)^2 \,,
\label{flat1} \\
\eta_V &\equiv& \frac{\mpl^2}{V} \frac{\partial^2V}{\partial\varphi^2}
\,,
\label{flat2}
\eea
which, together with the slow-roll approximation,
lead to the slow-roll behaviour
\be
3H\dot\varphi \approx - \frac{dV}{d\varphi}
\,.
\ee
Here $\mpl$ is the reduced Planck mass ($\mpl \equiv (8\pi G_N)^{-1}$).

Even without the slow-roll approximation, slices of uniform $\varphi$
correspond to comoving slices because a spatial gradient of $\varphi$
would give non-vanishing momentum density. 
Since comoving slices coincide with slices of uniform energy density,
the slices of uniform $\varphi$ coincide also with the latter.
 Also, since $\varphi$ is a Lorentz scalar,
its gauge transformation is the same as that of $\rho$. It follows \cite{vernizzi}
that
we can replace $\rho$ by $\varphi$ in the above expressions:
\bea
\zeta\one &=&  -\psi\one - \calh \frac{\delta\varphi\one}{\varphi_0'} \,, 
\label{gifirst2}\\
\zeta_2\su{MW} &=& -\psi_2^{\rm MW} - \calh \frac{\delta\varphi\two}{\varphi_0'}
+2\calh \frac{\delta\varphi\one}{\varphi_0'} \frac{\delta\varphi\one'}{\varphi_0'}
\nonumber \\
&&+ 2\frac{\delta\varphi\one}{\varphi_0'} \( \psi\one' + 2\calh \psi\one \)
\nonumber \\
&& - \( \calh \frac{\delta\varphi\one}{\varphi_0'} \)^2
\( \frac{\varphi_0''}{\calh \varphi_0'} - \frac{\calh'}{\calh^2} - 2 \)
\label{mwgi2}
\,, \\
\zeta_2 &=& -\psi_2 - \calh \frac{\delta\varphi_2}{\varphi_0'}
+2\calh \frac{\delta\varphi\one}{\varphi_0'} \frac{\delta\varphi\one'}{\varphi_0'}
+ 2\frac{\delta\varphi\one}{\varphi_0'} \psi\one' \nonumber \\
&&- \( \calh \frac{\delta\varphi_1}{\varphi_0'} \)^2
\( \frac{\varphi_0''}{\calh \varphi_0'} - \frac{\calh'}{\calh^2} \)
\label{zetagi}
\,.
\eea

\subsection{Multi-component inflation}

Now consider the case of multi-component inflation, where there is a family
of inequivalent inflationary trajectories lying in an $N$-dimensional
manifold of  field space. If the relevant part of the manifold is 
 not too big it will be a good approximation to take the fields to be 
canonically normalised. Then the inequivalent trajectories will be
curved in field space\footnote
{More generally
they will be 
non-geodesics, the 
geodesics being  the trajectories which 
 the background fields
could follow if there was 
 no potential term in the scalar Lagrangian \cite{rigopoulos}.}.
To define the trajectories 
one can choose a fixed basis in field space corresponding to 
fields $\phi_1,\cdots,\phi_N$.

Assuming canonical normalisation, multi-component slow-roll inflation is characterised
by the  conditions
\bea
\frac{\mpl^2}{2V^2} \left( \frac{\partial V}{\partial \phi_n} \right)^2 
\ll &1 \,,& \label{flat1a} \\
\frac{\mpl^2}{V} \left| \frac{\partial^2 V}{\partial \phi_n \partial\phi_m}
\right|  \ll &1 \,,& \label{flat2a} \\
3H\dot\phi_n \approx  &- \frac{\partial V}{\partial \phi_n} \,.& \label{svcon}
\eea

\begin{center}
\begin{figure}
\leavevmode
\hspace{2mm}
\hbox{\hspace{-1cm}%
\epsfxsize=2in
\epsffile{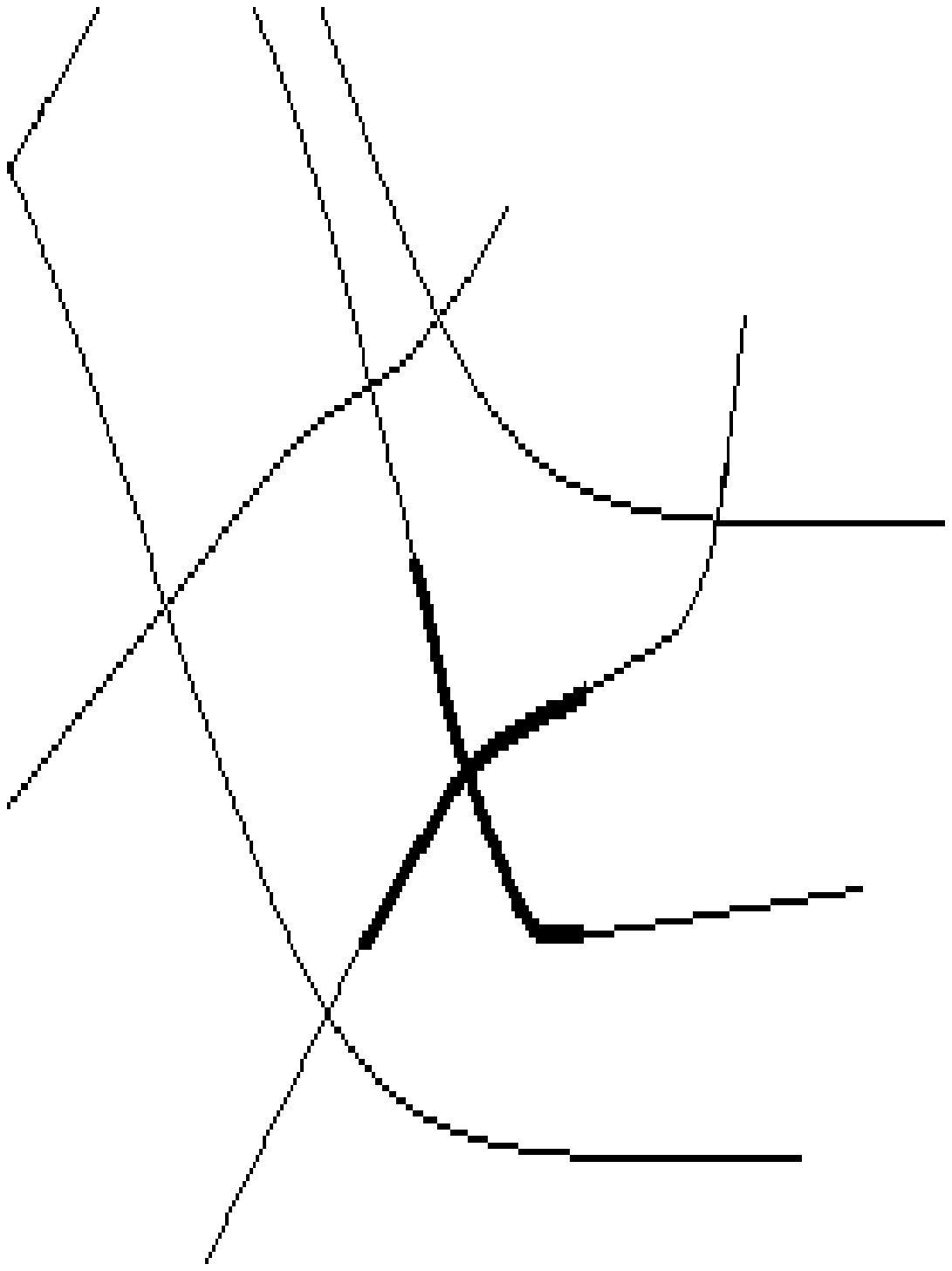}}
\hbox{\hspace{-1cm}%
\epsfxsize=2in
\epsffile{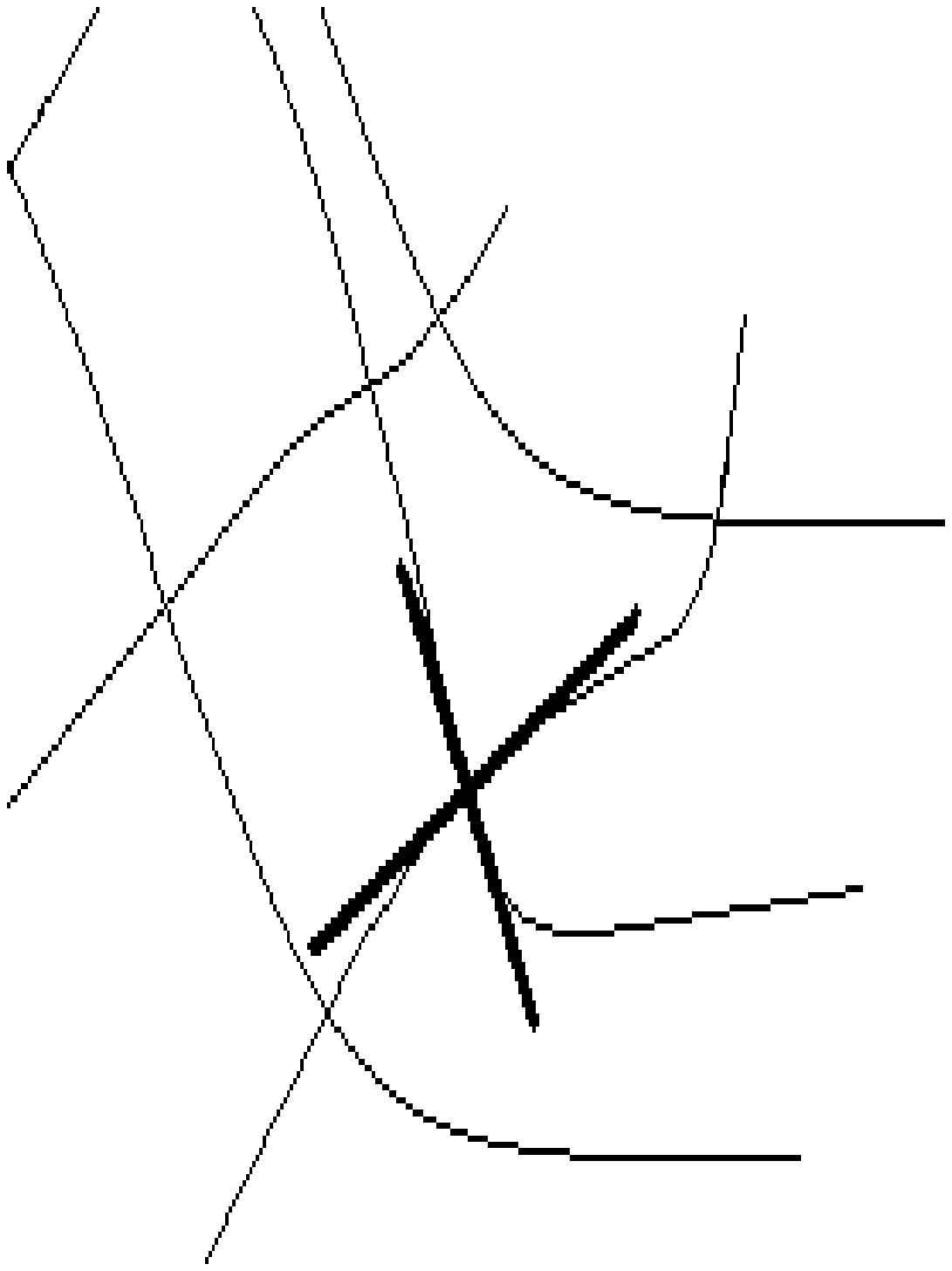}}
\put(-170,140){{\bf $\varphi$}}
\put(-215,163){{\bf $\sigma$}}
\put(-170,108){$V = {\rm const}$}
\put(-175,75){$V = {\rm const}$}
\put(-180,55){$V = {\rm const}$}
\put(-65,108){{\bf $\varphi$}}
\put(-65,68){{\bf $\sigma$}}
\put(-40,104){$V = {\rm const}$}
\put(-45,73){$V = {\rm const}$}
\put(-50,53){$V = {\rm const}$}
\put(-193,40){(a)}
\put(-75,40){(b)}
\vspace{-1.5cm}
\caption{Two different procedures 
for defining the fields in two-component
inflation. The fields are denoted by $\varphi$
and $\sigma$.
 (a) The field
$\varphi$  parameterises the 
distance along
the inflaton trajectories, with 
 uniform $\varphi$ corresponding to the equipotential
lines.
 The field $\sigma$ parameterises the 
distance along the equipotentials.
(b) The fields $\varphi$ and $\sigma$ are the components in
a fixed orthonormal basis, aligned with the inflationary trajectory
 at a certain point in field space. The value of 
$\varphi$ is now  the displacement 
 along the  tangent vector
and the value of  $\sigma$ is the displacement  along
the orthogonal vector. Working to second order in these
displacements, the equipotentials no longer coincide
with the lines of uniform $\varphi$. \label{fbasis}}
\end{figure}
\end{center}

The procedure of choosing a fixed basis 
is quite convenient for calculations, but a different
procedure leads to a perhaps simpler theoretical description.
This is to take $\varphi$ to parameterise the distance along the inflaton
trajectories, just as in single-component inflation, but now with the
proviso that uniform $\varphi$ corresponds to uniform field potential
(since we work in the slow-roll approximation, this means that the 
slices in field space of uniform $\varphi$ are orthogonal to 
the trajectories).
Then, in the slow-roll approximation, slices of spacetime with
uniform $\varphi$ will again coincide with slices of uniform density (see Fig.
\ref{fbasis}a).
Since $\varphi$ is a scalar, Eqs. (\ref{gifirst2}) and (\ref{mwgi2}) will then
be
valid. This is the simplest form of the gauge-invariant expression,
though for a practical calculation it may be better to write it in terms
of a fixed basis.

There is a subtlety here. For the first-order case we could
define $\varphi$ in a different way; around  a given point 
on  the unperturbed trajectory we could choose a fixed field
basis, with one of the basis vectors pointing along the trajectory,
and define $\varphi$ as the corresponding field component.
Then we could choose $\varphi$ to be canonically normalised in the 
vicinity of the chosen point in field space. 
That would not work at second order though,  because at that order
it makes a difference whether $\varphi$ is the appropriate parameterisation
of the  distance
along the trajectories (our adopted definition)
or the distance along a tangent vector to the trajectory (the 
alternative definition) (see Fig. \ref{fbasis}b). Only our adopted one will
make \eqs{mwgi2}{zetagi} valid.

\subsection{A third definition of the curvature perturbation}

The third definition in the literature 
applies only during inflation. It  was
given originally by Acquaviva {\it et. al.} \cite{acquaviva} for
the single-component case,
and the generalization to the multi-component case was noted by
Rigopoulos \cite{rigopoulos}. We shall denote this definition by
$\zeta\su{A}$.

The definition of Acquaviva {\it et. al.} and Rigopoulos is
\begin{eqnarray}
\zeta\su{A} _2 &=& -\psi_2^{\rm MW} - 
\mathcal{H} \frac{\delta\varphi\two}{\varphi_0'} \nonumber \\
&& - \frac{
(\psi_1' + 2\mathcal{H}\psi_1 
+ \mathcal{H}\delta\varphi\one' / \varphi'_0)^2}{
\mathcal{H}' + 2\mathcal{H}^2 - 
\mathcal{H}\varphi''_0 / \varphi'_0
}
\,. \label{abmr}
\end{eqnarray}
This is gauge-invariant by construction, with 
 $\varphi$ defined as in Figure 1(a).
 
It was pointed out by Vernizzi \cite{vernizzi} (actually in the context
of  single-component inflation)  
that comparing this  definition with \eq{mwgi0} gives simply
\be
\zeta_2\su{A} =
 \zeta_2\su{MW}
 - \frac{4 \mathcal{H}^2(\zeta_1)^2}{\mathcal{H}' 
+ 2 \mathcal{H}^2 - \mathcal{H}\varphi''/\varphi'}
\,.
\label{abmr2}
\ee
In the limit of slow-roll the denominator of the last term becomes
just $2\calh^2$, and then
\bea
\zeta_2\su{A} &=& \zeta_2
\,.
\eea
In other words, this third definition coincides with our adopted one
in the slow-roll limit.

Making use of the slow-roll parameters defined in Eqs. (\ref{epsilon}) and (\ref{eta}), the expression
in Eq. (\ref{abmr2}) gives to first-order in the slow-roll approximation
\be
\zeta_2\su{A} = \zeta_2 
 - (2\epsilon - \eta_V) (\zeta_1)^2 \label{vercom}
\,.
\ee

\section{The  evolution of the curvature perturbation}
\label{origin}

The simplest possibility for the origin of the observed
curvature perturbation  is that it comes from
the vacuum fluctuation of the inflaton field in a single-component model.
More recently other possibilities were recognised and we summarise the 
situation now. Although the  discussion is
usually applied to the magnitude of the curvature perturbation,
it applies equally to the non-gaussianity.

\subsection{Heavy, light and ultra-light fields}

On each scale the  initial epoch, as far as classical perturbations
are concerned, should be taken to be a few Hubble times after horizon
exit during inflation. The reason is that all such perturbations
 are  supposed to originate from
the vacuum fluctuation of one or more light 
scalar  fields, the fluctuation on each scale being 
promoted to a classical perturbation around the time of horizon exit.

Considering a fixed basis with canonical normalisation, a light field
is roughly speaking one  satisfying the 
flatness condition in \eq{flat2a}. 
The terminology is suggested by the important special case
that the effective potential during inflation is quadratic.
 Then, a light field is  roughly speaking 
that whose effective  mass during inflation is less than the value
 $H_*$ of the Hubble parameter. 
More precisely, the condition that the vacuum fluctuation be  promoted to 
a classical perturbation is
\cite{mijic}
\be
m < \frac{3}{2} H_* \label{lightfield}
\,.
\ee
{}From now on we focus on the quadratic potential, and take this as the
 the definition of a light field.
Conversely a heavy field may be  defined as one for which the 
condition in \eq{lightfield} is violated.

 During inflation light fields slowly roll according to \eq{svcon}
(with the vacuum fluctuation superimposed)
 while the heavy fields presumably are pinned down at an instantaneous
minimum of the effective potential. 
As we have seen, multi-component inflation takes place in a subspace of
field space. The fields in this subspace are light, but
their effective masses are  sufficient
to appreciably curve the inflationary trajectories.
In the case of both  multi-component and single-component inflation,
there could also be `ultra-light' fields, which do not appreciably curve
the inflationary trajectory and which therefore
have practically no effect on the dynamics of inflation.

\subsection{The evolution of the curvature perturbation}

To describe  the behaviour of  perturbations during the 
super-horizon era, without making too many detailed assumptions,
 it is useful to invoke the separate universe
hypothesis  \cite{bst,wands,lyth,tanaka} 
after smoothing on a given comoving scale much bigger than the horizon\footnote{
When considering linear equations, smoothing is equivalent to dropping
short wavelengths fourier components. In the nonlinear situation the smoothing
procedure could be in principle ambiguous. In a given situation one should state
explicitly which quantities are being smoothed.}.
According to this hypothesis the local evolution at each position is
that of some unperturbed universe (separate universe).
 Of course the separate universe hypothesis can and should be 
checked where there is a sufficiently detailed model.
However, it should be correct
on cosmological scales for a very simple reason. The unperturbed Universe
may be  defined as the one around us, smoothed on a scale
a bit bigger than the present Hubble distance.
In other words, the separate universe hypothesis is certainly valid
when applied to that scale. But the whole range of cosmological 
scales spans only a few decades.
This means that cosmological scales
are  likely to be huge compared with any
scale that is relevant in the early Universe, and accordingly that the 
separate universe hypothesis should be valid when applied to cosmological
scales even though it might fail on much smaller scales
(this expectation  was verified
in a preheating example \cite{ourreheat}
to which we  return later).

We are concerned with the curvature perturbation, which during the 
 super-horizon era 
 is conserved as long as the pressure is a unique function of the 
energy density (the adiabatic pressure condition). 
The 
 adiabatic pressure condition 
will be satisfied if and only if the separate universes are identical
(at least as far as the relation between pressure and energy density is
concerned) \footnote
{Of course the identity will only hold after making an 
 appropriate synchronization of the clocks at different positions. Having made that
synchronization, horizon entry will occur at different times
in different positions, which can be regarded as the origin
of the  curvature perturbation.}. 
The condition to have identical universes after a given epoch is  that
the specification of a  {\em single} quantity at that epoch is sufficient
to determine the entire subsequent evolution. 

In the case of single-component inflation,
the initial condition may be supplied by the local  value of the 
inflaton field, at the very 
beginning of the super-horizon era when it first becomes
classical.  Given the separate universe hypothesis, that is the only
possibility if the inflaton is the only light field ever to play a 
significant dynamical role. This means that {\em the curvature perturbation
generated at horizon exit during single-component inflation will be
equal to the one observed at the approach of horizon entry, provided
that the inflaton is the only light field ever to play a dynamical role.}

If inflation is multi-component, more than one field is 
by definition relevant during inflation. Then   the curvature perturbation
cannot be conserved during  inflation. The variation of the curvature 
perturbation during multi-component inflation is caused by the vacuum 
fluctuation orthogonal to the unperturbed inflationary trajectory, which
around the time of horizon exit
kicks the trajectory onto a nearby one so that the local trajectory
becomes position-dependent. After inflation is over, the curvature 
perturbation will be conserved if 
 the local trajectories lead to practically identical  universes.
In other words it will be conserved if the light (and ultra-light) fields,
orthogonal to the trajectory at the end of inflation, do not
affect the subsequent evolution of the Universe.

The curvature perturbation after inflation will vary if some
 light or  ultra-light field, orthogonal to the trajectory at the end of 
inflation, affects the subsequent evolution of the Universe (to be 
precise, affects  the pressure).
As we shall describe in Section \ref{after}, three types of scenarios have been proposed for
this post-inflationary variation of the curvature perturbation.

\section{Non-gaussianity} \label{gaussianity}

\subsection{Defining the non-gaussianity}

A gaussian perturbation is one whose Fourier components are uncorrelated.
All of its statistical properties are defined by its spectrum, and the 
spectrum  $\calp_g$ of a generic perturbation is conveniently 
 defined  \cite{treview,book}  by\footnote
{Technically the expectation values in this and the following expressions 
refer to an ensemble of universes but, because the stochastic properties 
of the
perturbations are supposed to be invariant under translations, the 
expectation
values can also be regarded as averages over the location of the observer
who defines the origin of coordinates.}
\be
\vev{ g(\bfk_1)g(\bfk_2) } = \frac{2\pi^2}{k^3} \delta^3(\bfk_1+\bfk_2) \calp_g(k)
\,,
\ee
the normalisation being chosen so that
\be
\vev{g^2(\bfx)} = \int^\infty_0 \calp_g(k) \frac{dk}{k}
\,.
\ee
On cosmological scales a  few Hubble times before 
horizon entry, observation shows that the
curvature perturbation is almost Gaussian with $\calp_\zeta^{1/2}
\simeq 10^{-5}$. 

The simplest kind of non-gaussianity 
that the  curvature perturbation 
could possess is  of the form
\bea
\zeta(\bfx) = \zeta\sub g(\bfx)  - \frac{3}{5} f\sub{NL} \( \zeta\sub g^2(\bfx)
 - \overline{\zeta\sub g^2} \)
\label{fnldef}
\,,
\eea
where
$\zeta\sub g$ is Gaussian with $\vev{\zeta\sub g}=0$,
and the non-linearity parameter $f\sub{NL}$ 
is independent of position.  We will call this {\em correlated $\chi^2$ non-gaussianity}.
Note that this definition assumes that $\vev\zeta
=0$, which means that the zero Fourier mode (spatial average) is 
dropped.

Following Maldacena \cite{maldacena}, we have inserted the prefactor
$-(3/5)$ so that in first-order perturbation theory our definition agrees
with that of Komatsu and Spergel \cite{spergel}, which is generally
the definition people use
when comparing theory with observation. Working in first-order
perturbation theory, these authors write  $\Phi(\bfx) = \Phi_g(\bfx) +
\fnl \left(\Phi_g^2(\bfx) - \overline{\Phi_g^2}\right)$, and  their
$\Phi$ is equal to $-3/5$ times our $\zeta$. 

One of the most powerful observational signatures of non-gaussianity
is a nonzero value for the three-point correlator, specified by the 
bispectrum $B$ defined by \cite{spergel,review}
\be
\vev{\zeta(\bfk_1)\zeta(\bfk_2)\zeta(\bfk_3)} = (2\pi)^{-3/2} B({k_1,k_2,k_3}) 
\delta^3(\bfk_1
+\bfk_2 + \bfk_3 ) 
\label{bdef}
\,.
\ee
For correlated $\chi^2$  non-gaussianity (with the gaussian term dominating)
\be
B({k_1,k_2,k_3}) = -\frac{6}{5}f\sub{NL}
\Big[ P_\zeta(k_1) P_\zeta(k_2)
+ {\rm \, cyclic \,} \Big]
\label{bchis}
\,,
\ee
where $P_\zeta(k) = 2\pi^2\mathcal{P}_\zeta(k)/k^3$. 
For any  kind of non-Gaussianity
one may use the above expression to define a function $f\sub{NL}
(k_1,k_2,k_3)$.

Given a calculation of
 $f\sub{NL}$ using first-order perturbation
theory, one expects  in general that going to second order will change
$f\sub{NL}$ by an amount of order 1. On this basis, one expects that a 
first-order calculation is  good enough if it yields
$|f\sub{NL}|\gg 1$, but that otherwise a second-order calculation will be
necessary. 

The definition \eq{bchis}  of $f\sub{NL}$ is  made using our adopted definition
of $\zeta$. If $\zeta$ in the definition is   replaced 
by $\zeta\su{MW}$ (with the zero Fourier mode dropped)
then $f\sub{NL}$ should be replaced by
\be
f\sub{NL}\su{MW}\equiv f\sub{NL} - \frac{5}{3} \label{minusone}\label{fnlmw}
\,.
\ee
To obtain this expression we used \eq{mwdef} and dropped terms higher than
second order. 

All of this assumes that the non-gaussian component of $\zeta$ is fully 
correlated with the gaussian component. An alternative possibility
\cite{bl}
that will be important for us is if $\zeta$ has the form
\bea
\zeta(\bfx) = \zeta_g (\bfx)  - \frac{3}{5} \tilde f\sub{NL} \( \zeta_\sigma^2(\bfx)
 - \overline{\zeta_\sigma^2} \)
\label{fnltildef}
\,,
\eea
where
$\zeta_g$ and $\zeta_\sigma$ are  uncorrelated  Gaussian perturbations,
normalised to have equal
spectra, and the  parameter $\tilde f\sub{NL}$ 
is independent of position.  We will call this {\em
uncorrelated $\chi^2$ non-gaussianity}.
It can be shown \cite{bl}
that in this case, $f\sub{NL}$ as defined by \eq{bchis} is given by
\be
f\sub{NL} \sim \( \frac{\tilde f\sub{NL}}{1300} \)^3
\,.
\ee

\subsection{Observational constraints on the non-gaussianity}

Taking $f\sub{NL}$ to denote the non-linearity parameter at the primordial 
era, let us consider the observational constraints.
Detailed calculations have so far been made  only with
 $f\sub{NL}$ independent of the wavenumbers, and only
by using  first-order perturbation theory for the evolution of the 
cosmological perturbations after horizon entry.
It is found  \cite{komatsu}
that present observation
 requires
 $|f\sub {NL}|\lsim 10^2$   
making the non-gaussian fraction at most of order $10^{-3}$. 
The use of first-order perturbation theory 
in this context  is amply justified.
Looking to the future though, it is found
that the
PLANCK satellite will either detect non-gaussianity or reduce the bound to
$|f\sub {NL}|\lsim 5$ \cite{spergel,review}, and that
foreseeable  future observations can reach
a level $|f\sub{NL}|\sim 3$ \cite{spergel,review}. 

Although the use of first-order perturbation
theory is not really justified for the latter estimates, we can safely
conclude that it will be  difficult for observation ever to detect
a value $|f\sub{NL}|\ll 1$. That is a pity because,  as we shall see,
such a value is predicted by some theoretical scenarios. On the other 
hand, other scenarios predict $|f\sub{NL}|$ roughly of order 1.
It will  therefore be of great interest to have detailed second-order
 calculations, to establish precisely the level of sensitivity that can
be achieved by future observations. A step in this direction has 
been taken in Refs. \cite{bartolotemp,creminelli}, where the large-scale 
cosmic microwave background anisotropy
is calculated to second order in terms of only the curvature perturbation 
(generalizing the first-order Sachs-Wolfe effect \cite{sachs}).

\section{The initial non-gaussianity}

\label{initial}

\subsection{Single-component inflation}

At first order, the curvature perturbation during single-component
inflation is Gaussian. Its time-independent spectrum is given by 
\cite{treview,book}
\be
\calp_\zeta (k) = \left[\left(\frac{H}{2\pi}\right)^2 \left(\frac{H}{\dot{\varphi}}\right)^2\right]_{k=aH}
\label{single}
\,,
\ee
and its  spectral index $n\equiv d\ln \calp_\zeta/d\ln k$ is given by
\be
n-1 = 2\eta_V -6\epsilon
\,.
\ee
The spectrum $r$ of the 
 tensor perturbation, defined 
 as a fraction of $\calp_\zeta$, is also given
in terms of the slow-roll parameter $\epsilon$:
\be
r = 16\epsilon.
\ee

{\em If the curvature perturbation does not evolve after single-component
inflation is over} observation constrains $n$ and $r$, and hence
the slow-roll parameters $\eta_V$ and $\epsilon$. A current bound
\cite{observation} is $-0.048<n-1<0.016$ and $r<0.46$. The second bound
gives $\epsilon<0.029$, but barring an 
accurate cancellation the first bound gives $\epsilon \lsim  0.003$.
In most inflation models $\epsilon$ is completely negligible and then 
the first bound gives $-0.024<\eta_V<0.008$ (irrespective of slow-roll inflation models,
the  upper bound in this expression  holds generally,
and the lower bound is badly violated only if there is an accurate 
cancellation). The bottom line of all this
 is that $\epsilon$ and $|\eta_V|$ are both constrained to be 
$\lsim 10^{-2}$.

Going to 
second order, Maldacena \cite{maldacena} has calculated the bispectrum
during single-component inflation (see also Refs. \cite{seery,rst,calcagni,sb2,mollerach,rs,rst1}). His
result may be written in the form
\be
f\sub{NL} = \frac{5}{12} \(2\eta_V - 6\epsilon -2 \epsilon f(k_1,k_2,k_3) \) 
\label{fnlmal}
\,,
\ee
with  $0 \leq f\leq 5/6$.
By virtue of the slow-roll conditions, $|f\sub{NL}|\ll 1$ \footnote
{Near a maximum of the potential `fast-roll' inflation \cite{fastroll,lotfi}
can take place with  $|\eta_V|$ somewhat
bigger than 1. Maldacena's calculation does not apply to that case but,
presumably, it gives initial non-gaussianity $|\fnl|\sim 1$. However, the
precise initial value of $\fnl$ in this case is not important because the 
corresponding initial 
spectral index is far from $1$, which means that the initial
curvature perturbation must be negligible.}.
In other words, the curvature perturbation $\zeta$, 
{\em as we have defined it},
is almost Gaussian during single-component inflation.

From \eq{vercom}
$\zeta\su{A}$ is also practically gaussian, but this quantity is defined
only during inflation and therefore could not be considered as a replacement
for $\zeta$. More importantly, $\zeta\su{MW}$ has significant 
non-gaussianity because,  from
\eq{minusone}, it corresponds to   $f\sub{NL}\su{MW} \approx -5/3$.  

One may ask why it is our $\zeta$ and not $\zeta\su{MW}$ which is gaussian
in the slow-roll limit\footnote{We
thank Paolo Creminelli for enlightening correspondence on this question.}. 
 One feature that distinguishes our $\zeta$,
is that any part of it can be absorbed into the scale factor without altering
the rest; indeed
\be
g_{ij}=\delta_{ij}a^2(\eta)e^{2\zeta_1+\zeta_2}
=\delta_{ij}\tilde a^2(\eta)e^{\zeta_2}
\,,
\ee
with $\tilde a = ae^{\zeta_1}$ (if we tried to do that with $\zeta\sub{MW}$,
the part of $\zeta$ not absorbed would have to be re-scaled). This means that
an extremely long-wavelength and possible large part of $\zeta$ has no local
significance. It also means, in the context of perturbation theory, 
  that the first-order part of $\zeta$ can be absorbed into the scale factor
when discussing the second-order part. 
However, the gaussianity of $\zeta$ does not seem to be 
related directly to this feature.
Rather, it has to do with  the gauge
transformation, relating quantities $\psi_A$ and $\psi_B$ defined on 
different slicings.

With our definition \cite{sasaki1}, the gauge transformation is
\be
\psi_A(t,\bfx)-\psi_B(t,\bfx)  = - \Delta N_{AB}(t,\bfx)
\label{psigt}
,
\ee
where $\Delta N_{AB}$ is the number of $e$-folds of  expansion
going from a slice $B$ to a slice $A$, both of them corresponding to 
time $t$ \footnote
{This expression
is valid even when the tensor perturbation is included \cite{sasaki1}. As a result,
 the gauge-invariant expressions mentioned earlier are still valid
in that case, as are the results based on them including the present
discussion. \label{teninc}}.
In writing this expression  we used
 physical time $t$ instead of conformal time,
the two related by $dt=a d\eta$. 
Along a comoving worldline, the number of $e$-folds of expansion is defined as
$N\equiv \int \tilde H d\tau$
where $\tilde H$ is the local Hubble parameter and $d\tau$ is the proper
time interval \cite{sasaki1}.

To understand the relevance of this result, take $\psi_B=0$ 
and $\psi_A=-\zeta$.
The pressure is adiabatic during single-component inflation, 
which means that $dt$ can be identified with the proper time
interval $d\tau$, and the proper expansion rate on slicing $A$
is uniform \cite{sasaki1}. As a result, to second order, 
\bea
\zeta  &=& H(t) \Delta t(t,\bfx) + 
\frac12 \dot H(t) \(\Delta t(t,\bfx) \)^2 \nonumber \\
&\simeq &  H \Delta t(t,\bfx) +
\frac12 \frac{\dot H}{H^2} \( H \Delta t(t,\bfx) \)^2 \nonumber \\
&\simeq &  H \Delta t(t,\bfx) 
\label{gt1}
\,.
\eea
In the last line we made the slow-roll approximation, and from the second
line we can see that the error in $f\sub{NL}$  caused by this approximation
 is precisely $\epsilon$.

We  also need the gauge transformation for the inflaton
field $\varphi$ in terms of $\Delta t$. 
Since the slices correspond to the same coordinate
time,  the unperturbed inflaton 
field can be taken to be the same on each of them which means that the gauge
transformation for $\delta \varphi$ is
\be
\delta\varphi_A(t,\bfx) - \delta\varphi_B(t,\bfx) = 
\Delta \varphi_{AB} (t,\bfx)
,
\ee
where $\Delta\varphi_{AB}$ is the change in $\varphi$ going from slice
$B$ to slice $A$. But slice $A$ corresponds to uniform $\varphi$,
which means that on slice $B$ to second order
\bea
H(t)\frac{\delta\varphi_B(t,\bfx)}{\dot \varphi_0}  &=& -H(t) \Delta t (t, \bfx) -
\frac12 H(t) \frac{\ddot \varphi_0}{\dot \varphi_0} (\Delta t (t, \bfx))^2  \nonumber \\
& \simeq & -H \Delta t (t, \bfx) -
\frac12 \frac{\ddot \varphi_0}{H \dot \varphi_0} (H \Delta t (t, \bfx))^2 \nonumber \\
& \simeq  &  -H \Delta t (t, \bfx)
\label{gt2}
\,,
\eea
where in the last line we used the slow-roll approximation. We can see that the
fractional 
error caused by this approximation is $\ddot\varphi_0/H\dot\varphi_0
=\epsilon -\eta_V$.

Combining \eqs{gt1}{gt2} we have in the slow-roll approximation
\be
\zeta \simeq  -H(t) \frac{\delta \varphi_B (t, \bfx)}{\dot \varphi_0}
\,,
\ee
with fractional error of order $\max\{\eta_V,\epsilon\}$ (this can also be seen directly
from \eqs{gifirst2}{zetagi} evaluated
with \mbox{$\psi=0$}, but we give the above argument because it explains why
the result is valid for $\zeta$ as opposed to $\zeta\su{MW}$).

The final and crucial step is to observe that 
 in the slow-roll approximation $\varphi_B$ is  gaussian,
with again a fractional error of order
 $\max\{\eta_V,\epsilon\}$. This was demonstrated by Maldacena
\cite{maldacena} but the basic reason is very simple. The  non-gaussianity
of $\varphi$ comes either from  third and higher derivatives of $V$
(through the field equation in unperturbed spacetime) or else through
the back-reaction (the perturbation of spacetime); but
the  first effect is small \cite{treview,book} by virtue of the flatness
requirements on the potential,
and the second effect is small because $\dot\varphi_0/H^2$ is small \cite{book}.
This explains why  $\zeta$ with our adopted definition is practically
Gaussian by virtue of the slow-roll approximation.

\subsection{Multi-component inflation}

The flatness and slow-roll conditions \eqss{flat1a}{flat2a}{svcon} 
ensure that 
 the curvature of the inflationary trajectories 
is small during the few Hubble times around horizon exit, during which
the quantum fluctuation is promoted to a classical perturbation.
As a result,  the  {\em initial} curvature perturbation 
in first-order perturbation theory,
is still given by \eq{single} in terms of the field $\varphi$
that we defined earlier.

What about the initial non-gaussianity generated at second order?
In the approximation that  the  curvature of the trajectories
around horizon exit is  completely negligible, the orthogonal fields
are strictly massless. Such fields would not
affect  Maldacena's second-order calculation, which would therefore
still give the initial non-gaussianity.
It is not quite clear
whether the curvature can really be neglected, as it can be in the 
first-order case, and it may therefore be that the initial non-gaussianity
in multi-field inflation is different from Maldacena's result.
Even if that is the case though, we can safely say that the initial 
non-gaussianity corresponds to $|f\sub{NL}|\ll 1$, since the curvature
of the trajectories is certainly small.

\section{The evolution after horizon exit}
\label{after}

\subsection{Single-component inflation and  $\zeta\su{A}_2$}

During single-component inflation the curvature perturbation
$\zeta$, as we have defined it, does not evolve. From its definition
\eq{mwdef}, the same is true of 
 $\zeta_2\su{MW}$.

 In contrast
$\zeta_2\su{A}$, given by Eq. (\ref{vercom}),
will have the slow variation \cite{vernizzi}
\be
\dot \zeta_2\su{A} \approx -(2\dot\epsilon - \dot\eta_V) (\zeta\one)^2
\label{dotzetar}
\,.
\ee
This variation has no physical significance, being an artifact of the
definition.

Using a particular   gauge,
Acquaviva {\it et. al.} \cite{acquaviva}
have calculated $\dot\zeta_2\su{A}$ in terms of first-order
quantities $\psi\one$, $\delta\varphi\one$, and their derivatives,
and they have displayed the
result as an indefinite integral
\be
\zeta_2\su{A}(t) = \int^t A(t) dt + B(t)
\label{aeq}
\,.
\ee
Inserting an initial condition, valid a few Hubble times after horizon exit, this becomes
\be
\zeta_2\su{A}(t) = \zeta_2\su{A}(t\sub i) + \int^t_{t\sub i} A(t) dt +
\left. B \right|^t_{t\sub i}
\,. \label{time}
\ee
In view of our discussion, it is clear that these equations will, if
correctly
evaluated, just reproduce the time dependence of \eq{dotzetar}.

The authors of Ref. \cite{acquaviva} also present an equation
for $\dot\zeta_2\su{A}$, again involving only first-order quantities,
 which is valid also before horizon entry. Contrary to the claim of the 
authors, this   classical
equation  cannot by itself be used to calculate the
initial value (more precisely, the
stochastic properties of the initial value) of $\zeta_2 \su{A}$.
In particular, 
it cannot by itself  reproduce Maldacena's calculation of the 
bispectrum.

It is true of course that in the Heisenberg picture the quantum operators
satisfy the classical field equations. In  first-order perturbation
theory,  where the equations
are linear, this allows one to calculate the curvature perturbation
without going to the trouble of calculating the second-order action
\cite{book} (at the $n$th order of perturbation theory 
the action has to be evaluated to order $n+1$ if it is to be used).
At second order in perturbation theory it remains to be seen whether the 
Heisenberg picture  can provide a useful alternative to Maldacena's 
calculation, who adopted the interaction  picture and calculated the 
action to third order. 

\subsection{Multi-component inflation}

During multi-component inflation the curvature perturbation by definition varies
significantly along a generic trajectory, 
which means that non-gaussianity is generated at some level.
So far only a limited range of models
has been investigated 
\cite{multiinfng1,multiinfng2,multiinfng3,multiinfng4,enqvist,enqvistnew}.
To keep the spectral tilt within observational bounds, the unperturbed
trajectory in these  models has to be specially chosen, but the
choice might  be justified by a suitable initial condition.

We shall consider here  a  calculation  by
Enqvist and V$\ddot{{\rm a}}$ihk$\ddot{{\rm o}}$nen in Ref. \cite{enqvist}.
Following the same line as Acquaviva {\it et. al.} \cite{acquaviva}, 
they study a two-component inflation model, in which the only 
important parts of the potential are 
\be
V(\varphi,\sigma) = V_0 +\frac12 m_\sigma^2 \sigma^2 + 
 \frac{1}{2}m_\varphi^2 \varphi^2 
\label{effv}
\,.
\end{equation}
The masses are both supposed to be less than $(3/2)H_*$, so that this is a
two-component inflation model, and the above form of the potential is
supposed to hold for some number $\Delta N$ of $e$-folds after cosmological
scales leave the horizon. They
take the unperturbed inflation trajectory to have $\sigma_0=0$,
and the idea is to calculate the amount of non-gaussianity generated
after $\Delta N$ $e$-folds. Irrespective of any  later evolution,
this calculated non-gaussianity
will represent the minimal observed one (unless 
non-gaussianity generated later happens to cancel it).

It is supposed  
that the condition $\sigma_0=0$, as well as the ending of inflation,
will come from a tree-level hybrid potential,
\begin{equation}
V(\varphi,\sigma) = V_0 -\frac{1}{2}m_0^2 \sigma^2 + \frac{1}{4}\lambda \sigma^4 + \frac{1}{2}m_\varphi^2 \varphi^2 + \frac{1}{2}g^2 \sigma^2 \varphi^2
\label{fullv}
\,.
\end{equation}
Like the original authors though, we shall not investigate the extent to which 
 \eq{fullv} can reproduce \eq{effv} for at least some number of
$e$-folds. We just focus on \eq{effv}, with the assumption $\sigma_0=0$
for the unperturbed trajectory. 

Because $\sigma_0=0$, the unperturbed trajectory is straight, and
at first order the curvature perturbation $\zeta$ is conserved.
This is not the case though at second order.
Adopting the definition $\zeta\su{A}$, the authors of  \cite{enqvist}
give  an expression for $\zeta\su{A}_2$ similar to that in Eq. (\ref{time}) 
describing the evolution of the second-order curvature perturbation on 
superhorizon scales\footnote
{The fields $\varphi$ and $\sigma$ in \eq{effv} are supposed to be canonically
normalised, which means that $\varphi$ is {\em not} the field appearing in 
the Rigopoulos definition \eq{abmr} of $\zeta\su{A}$. Instead the authors
of \cite{enqvist} give an equivalent definition in terms of the canonically
normalised fields.}.
This equation, 
in the generalized longitudinal gauge,
 reads (from Eq. (67) in Ref. \cite{enqvist}):
\begin{eqnarray}
&&\zeta_2\su{A}(t) - \zeta_2\su{A}(t\sub i) = \nonumber \\
&&-\frac{1}{\epsilon H \mpl^2} \Big \{ \int^{t}_{t\sub i} \Big[6H \Delta^{-2} 
\partial_i (\delta\dot{\sigma}\one \partial^i \delta\sigma\one) - 
2(\delta\dot{\sigma}\one)^2 \nonumber \\
&&+ 4 \Delta^{-2} \partial_i (\delta\dot{\sigma}\one 
\partial^i \delta\sigma\one)^\cdot + m_\sigma^2 (\delta\sigma\one)^2 
\nonumber \\
&&+ (\epsilon - \eta_V)6H\Delta^{-4}\partial_i(\partial_k \partial^k 
\delta\sigma\one \partial^i \delta\sigma\one)^\cdot \nonumber \\
&&+ (\epsilon - \eta_V)H\Delta^{-4}\partial_i\partial^i(\partial_k \delta
\sigma\one \partial^k \delta\sigma\one)^\cdot \nonumber \\
&&- 3\Delta^{-4}\partial_i(\partial_k \partial^k \delta\sigma\one 
\partial^i \delta\sigma\one)^{\cdot\cdot} \nonumber \\
&&- \frac12 \Delta^{-4}\partial_i\partial^i(\partial_k \delta\sigma\one 
\partial^k \delta\sigma\one)^{\cdot\cdot}\Big] dt \nonumber \\
&&+ \Big[ - \Delta^{-2}\partial_i(\delta\dot{\sigma}\one \partial^i 
\delta\sigma\one) + 3\Delta^{-4}\partial_i(\partial_k \partial^k 
\delta\sigma\one \partial^i \delta\sigma\one)^\cdot \nonumber \\
&&+ \frac12 \Delta^{-4}\partial_i\partial^i(\partial_k \delta\sigma\one 
\partial^k \delta\sigma\one)^\cdot \nonumber \\
&&+ 3\epsilon H\Delta^{-4}\partial_i(\partial_k \partial^k \delta
\sigma\one \partial^i \delta\sigma\one) \nonumber \\
&&+ \frac{\epsilon H}{2}\Delta^{-4}\partial_i\partial^i(\partial_k 
\delta\sigma\one \partial^k \delta\sigma\one) \Big] \Big|^{t}_{t\sub i} 
\Big \} \,, \label{ourev} 
\end{eqnarray}
where $\Delta^{-2}$ is the inverse of the Laplacian operator.

Assuming that this expression is correct, we consider the non-gaussianity
it may generate. Following the original authors, we note first that
\begin{equation}
\left| H \frac{\delta\sigma\one}{\dot \varphi_0} \right| \sim 
\left| H \frac{\delta\varphi\one}{\dot \varphi_0} \right| = 
\left| \zeta_1 \right| \equiv {\rm constant},
\end{equation}
which is a good approximation since the first-order perturbation equation of the 
effectively massless field $\sigma$ is the same as the first-order 
perturbation equation of the inflaton field $\varphi$ on superhorizon scales. 
Moreover, the time derivative of the first-order perturbation in $\sigma$ 
can be estimated as
\begin{equation}
\left| \delta \dot \sigma\one \right| \sim \frac{m_\sigma^2}{H} \left| 
\delta\sigma\one \right|,
\end{equation}
assuming slow-roll conditions.
If we also assume that $H$ and $m_\sigma^2$ are almost constants in time, 
we end up with
\begin{eqnarray}
\zeta_2\su{A}(t) - \zeta_2\su{A}(t\sub i) &=& -\frac{1}{\epsilon H \mpl^2} 
\int^{t}_{t\sub i} \Big[ 6H \Delta^{-2} \partial_i (\delta\dot{\sigma}\one 
\partial^i \delta\sigma\one) \nonumber \\
&& - 2(\delta\dot{\sigma}\one)^2 + m_\sigma^2 (\delta\sigma\one)^2 \Big] 
dt \label{our3}
\,.
\eea

For $m^2/H^2\ll 1$ the first and third term in the integrand dominate, whereas
all the three terms become of the same order of magnitude for $m^2/H^2\sim 1$ which is the limit of
applicability of the calculation. In any case, the typical magnitude of the right-hand
side is of order
\be
 \Delta N \frac{m_\sigma^2}{H^2}
 \left|\zeta_1\right|^2 \,, \label{our2}
\ee
with $\Delta N$ the number of e-folds specified by the integral limits. 
This looks big, but we have to remember
that  the right hand side
is uncorrelated with the inflaton perturbation $\delta\phi$ which generates 
$\zeta_1\su A$. Therefore, \eq{fnltildef} as opposed to \eq{fnldef} 
applies, and 
 we would need $\Delta N\sim 1300$ to get even  $f\sub{NL}\sim 1$, which is
impossible.

These conclusions differ sharply from those of
 Enqvist and V$\ddot{{\rm a}}$ihk$\ddot{{\rm o}}$nen \cite{enqvist}
who actually find $\zeta_2^A \propto \mathcal{O} (\epsilon, \eta_V, m_\sigma^2 / H^2)%
(\zeta_1)^2$. \footnote{In the proper treatment of the integral and its initial condition 
the factors $\epsilon (\zeta_1)^2, 
\eta_V (\zeta_1)^2$, and $m_\sigma^2 /
H^2 (\zeta_1)^2$ cancel out since the evaluated quantity is in this case
$\zeta_2\su{A}(t) - \zeta_2\su{A}(t\sub i)$, and $\zeta_1$ is conserved.
This was not taken into account in Ref. \cite{enqvist}.}
Their estimate of the right-hand side of \eq{our3} does not contain our 
factor 
$\Delta N$, but much more importantly they estimate $f\sub{NL}$ as if
the right hand side were fully correlated with $\zeta_1\su A$ to conclude
that the model can give $f\sub{NL}\sim 1$.

The reason of the discrepancy lies in the Eq. (71) in Ref. \cite{enqvist},
which we write in the same form as our Eq. (\ref{our2}):
\begin{eqnarray}
\zeta_2\su{A}(t) - \zeta_2\su{A}(t\sub i) &=& - \frac{1}{\epsilon H \mpl^2}
\int^{t}_{t\sub i} \Big[ 6H \Delta^{-2} \partial_i (\delta\dot{\sigma}\one
\partial^i \delta\sigma\one) \nonumber \\
&& -2(\delta\dot{\sigma}\one)^2 + m_\sigma^2 (\delta\sigma\one)^2 \Big] dt
\nonumber \\
&=& -\frac{1}{\epsilon H \mpl^2} \int^{t}_{t\sub i} \Big[-2 \Delta^{-2}
\partial_i (\delta\ddot{\sigma}\one \partial^i \delta\sigma\one) \nonumber
\\
&& -2(\delta \dot{\sigma}\one)^2\Big], \label{vaih_main_2}
\end{eqnarray}
where the equation of motion $\delta\ddot{\sigma}\one + 3H
\delta\dot{\sigma}\one + m_\sigma^2 \delta\sigma\one = 0$ has been used.
Enqvist and V$\ddot{{\rm a}}$ihk$\ddot{{\rm o}}$nen seem to have neglected the
factor $\delta\ddot{\sigma}\one$ in the above expression, keeping only the
term $-2(\delta \dot{\sigma}\one)^2$ in the integrand, which gives
the partial estimate
\be
\zeta_2\su{A}(t) - \zeta_2\su{A}(t\sub i) \sim \Delta N \frac{m_\sigma^4}{H^4}
\left|\zeta_1\right|^2 << \frac{m_\sigma^2}{H^2}
\left|\zeta_1\right|^2.
\ee
This last step is wrong because $\delta
\ddot{\sigma}\one$ can only be neglected compared with $3H \delta
\dot{\sigma}\one
+ m_\sigma^2 \delta \sigma\one$, which is not present in the integrand in Eq.
(\ref{vaih_main_2}). Moreover, what we need is the integral of $\delta
\ddot{\sigma}\one$, i.e.
$\delta \dot{\sigma}\one$, which is in any case non-negligible.  That is the
origin of our new term
\be
\Delta N \frac{m_\sigma^2}{H^2} \left|\zeta_1\right|^2,
\ee
in Eq. (\ref{our2}). As a conclusion the level of non-gaussianity in this
hybrid-type model seems to be much smaller than previously thought as
the condition $m_\sigma \lsim H$ is presumably not
satisfied for a sufficient number of e-folds.

\subsection{Preheating}

Now we turn to the possibility that significant non-gaussianity could be
generated during preheating. Preheating is the term used to describe
the energy loss by scalar fields which might occur
between the end of inflation and reheating \cite{preheat1,kofman}, 
the latter being taken to 
correspond to the decay of individual particles which leads
to more or less  complete thermalisation of the Universe. Preheating 
typically produces marginally-relativistic particles, which 
decay before reheating.

It was suggested a long time ago \cite{bkm1,bkm2} 
that preheating might cause
the cosmological curvature perturbation to vary at the level of first-order 
perturbation theory, perhaps providing its  main origin.
More recently it has been suggested
\cite{enqvistnew,ev1,ev2}  that
 preheating might
cause the curvature perturbation to vary at second order, 
providing the main source of its
 non-gaussianity. 

If  the separate universe hypothesis is correct, a variation of the 
curvature perturbation during preheating can occur only in models
of preheating which contain a non-inflaton field that  is light during
inflation. This is not the case for the usual  preheating models
that were considered in \cite{enqvistnew,ev1,ev2},
and accordingly one does not expect that significant non-gaussianity will
be generated in those models\footnote
{The preheating model considered in \cite{enqvistnew} contains a field
which may be heavy or light; we refer here to the part of the 
calculation that considers the former case.}.
 This is not in conflict with the findings
of \cite{enqvistnew,ev1,ev2} 
because the  curvature perturbation is not actually
considered there.
Instead the perturbation $\psi^{\rm MW}$ in the longitudinal
gauge is considered, which is only indirectly related to $\zeta$
by Eqs. (\ref{mwdef}), (\ref{psigi}) and (\ref{mwgi0}) \footnote
{The slices
of the longitudinal gauge are orthogonal to the threads of zero shear,  and 
 $\psi^{\rm MW}$ on them is very different from
the curvature perturbation
 $\zeta$.}.
We conjecture that non-gaussianity for 
the curvature perturbation on cosmological scales
is not generated in the usual preheating
models, but that  instead the curvature perturbation  remains constant
on cosmological scales. This should of course be checked, 
in the same spirit that the constancy of the curvature perturbation 
was checked at the first-order level \cite{ourreheat}.

The situation is different for preheating models which 
contain a non-inflaton field that  is light during
inflation. At least three types of models have been proposed 
with that feature \cite{steve1,steve2,kr,enqvistnew}.
Except for \cite{enqvistnew} only the magnitude of the curvature 
perturbation has been considered, but in all three cases it might be
that  significant non-gaussianity is also generated. 

\subsection{The  curvaton scenario}

In the simplest version of
 the curvaton scenario \cite{curvaton1,curvaton2},
the  curvaton field $\sigma$ 
is ultra-light during inflation and has no significant
evolution until
 it starts to oscillate during some 
 radiation-dominated era. Until this oscillation gets under way,
 the
curvature perturbation is supposed to be negligible
 (compared with its final observed value).
 The potential during the oscillation
is taken to be quadratic, which will be
a good approximation after a few Hubble times even if it fails initially.
The curvature perturbation 
is generated during the oscillation, and is supposed to be conserved
after the curvaton decays. Here we give a generally-valid formula
for the non-gaussianity in the curvaton scenario, extending somewhat 
the earlier calculations.

The local energy density $\rho_\sigma$ of the  curvaton field
is given by
\begin{equation}
\rho_\sigma(\eta,{\bf x}) \approx \frac{1}{2} m_\sigma^2 \sigma_a^2(\eta,{\bf x}) \,, \label{ed}
\end{equation} 
where $\sigma_a(\eta,{\bf x})$ represents the amplitude of the oscillations
and $m_\sigma$ is the effective mass.
It is proportional to $a(\eta, {\bf x})^{-3}$ where $a$ is the locally-defined scale
factor. This means that the perturbation $\delta\rho_\sigma/\rho_\sigma$
is conserved if the slicing is chosen so that the expansion going from
one slicing to the next is uniform \cite{lyth}.
 The flat slicing corresponding
to $\psi^{\rm MW}=0$ has this property \cite{lyth,sasaki1} and accordingly 
$\delta\rho_\sigma$
is defined on that slicing.

Assuming that the fractional perturbation is small (which we shall see is
demanded by observation) it is given by
\be
\frac{\delta\rho_\sigma}{\rho_\sigma} = 2\frac{\delta\sigma_a}{\sigma_a}
+ \( \frac{\delta\sigma_a}{\sigma_a} \)^2
\label{delrhos}
\,.
\ee
We first assume that 
$\sigma(\bfx)$ has no evolution between 
inflation and
the onset of oscillation. Then
$\delta\sigma_a/\sigma_a$ will be equal to its  value just after
horizon exit, which we saw earlier will be practically gaussian.

The total density perturbation is given by
\be
\frac{\delta\rho}{\rho} = \Omega_\sigma \frac{\delta\rho_\sigma}{\rho_\sigma}
\,,
\ee
where $\Omega_\sigma\equiv \rho_\sigma/\rho\propto a$ 
is the fraction of energy density contributed by the curvaton.
Adopting the sudden-decay
approximation, the constant curvature  perturbation obtaining after 
the curvaton decays is given by  Eqs. (\ref{psigi}) and (\ref{zetatwo}), 
evaluated just
before curvaton decay and with $\psi=0$. In performing that calculation,
the exact expression \eq{delrhos} can, without loss of generality, be
identified with the first-order part 
$\delta\rho_{\sigma_1}/\rho_{\sigma_0}$, the second- and higher-order
parts being set at zero.

Adopting the first-order curvature perturbation in \eq{psigi},   one finds
 \cite{curvaton2}  chi-squared non-gaussianity
coming from the second term of \eq{delrhos},  
\be
f\sub{NL}= - \frac{5}{4r} \,,
\ee
with 
\be
r\equiv \frac{3\rho_\sigma}{4\rho_r + 3\rho_\sigma}
\label{rlin}
\,,
\ee
evaluated just before decay ($\rho_r$ is the radiation density).
Going to the second-order expression one finds 
\cite{curvaton}
additional chi-squared
non-gaussianity. The final non-linearity parameter
$f\sub{NL}=f\sub{NL}\su{MW}+5/3$ is given by
\begin{equation}
f_{NL} = \frac{5}{3} + \frac{5}{6}r - \frac{5}{4r}
\,,
\end{equation}

If $\Omega_\sigma\ll 1$ then $f\sub{NL}$ is strongly negative and the 
present bound on it requires $\Omega_\sigma \gsim 0.01$ (combined
with the typical value $\zeta\sim 10^{-5}$, 
this requires  $\delta \rho_\sigma/\rho_\sigma\ll 1$
as advertised). If instead $\Omega_\sigma=1$ to good accuracy,
then $f\sub{NL}= + 5/4$. Either of these possibilities may
be regarded as generic whereas the intermediate possibility 
($|f\sub{NL}|\sim 1$ but $f\sub{NL}\neq 5/4$) requires a special 
value of $\Omega_\sigma$ just a bit less  than $1$.

Finally, we consider  the case that $\sigma$ evolves 
between horizon exit and the era when the sinusoidal oscillation begins.
If $\sigma_a$ (the amplitude of  oscillation at the latter era)
is some function $g(\sigma_*)$ of the value  a few Hubble times after
horizon exit, then
\be
\delta\sigma_a = g' \delta \sigma_* + \frac12 g'' (\delta\sigma_*)^2
\,,
\ee
where the prime means derivative with respect to $\sigma_*$.
Repeating the above calculation one finds
\be
f\sub{NL} = \frac53 + \frac56 r -\frac5{4r}
\( 1 + \frac{g g''}{g'^2} \)
\,.
\ee
The final term  is the first-order result
(given originally in \cite{lowscale}),
the middle term
is the second-order correction
found in \cite{curvaton}, and the first
term converts from $f\sub{NL}\su{MW}$ to $f\sub{NL}$.

\subsection{The inhomogeneous reheating scenario}

The final scenario  that has been suggested for the origin of the 
curvature perturbation
is its generation during some  spatially inhomogeneous reheating process
\cite{inhomog1,inhomog2,inhomog3,inhomog4}.
Before  a reheating process the cosmic fluid is dominated by matter 
(non-relativistic  particles, 
or small scalar field oscillations which are equivalent to particles)
which then decay into thermalised radiation. At least one reheating 
process, presumably, has to occur to give the initial condition for
Big Bang Nucleosynthesis, but there might be more than one.

The inhomogeneous reheating scenario in its simplest form supposes that the curvature 
perturbation is negligible before the relevant reheating process, and
constant afterwards. The inhomogeneous reheating 
corresponds to a spatially varying value (a perturbation)
of the local Hubble parameter
$H\sub{reh}(\bfx)$ at the decay epoch
(or equivalently of the local energy density), and this
generates the final curvature perturbation. The perturbation in
$H\sub{reh}$ occurs presumably  because
it  depends on the value of some non-inflaton `modulon'
field $\chi$ which was light or ultra-light during inflation.

In contrast with the curvaton scenario, where the form
$\rho_\sigma$ can reasonably be taken as $\rho_\sigma\propto
\sigma^2$,  the inhomogeneous reheating scenario does not suggest any 
particular form for  $H\sub{reh}(\chi)$. Depending on the form,
the inhomogeneous reheating scenario presumably can produce a wide range of 
values for $\fnl$.
 
\section{Conclusions} \label{conclusions}

We have examined a number of scenarios for the production of 
a non-gaussian primordial curvature perturbation,
presenting the results with a unified notation.  These are the single-component inflation, multi-component inflation,
preheating, curvaton, and inhomogeneous reheating scenarios. 
Although the trispectrum may give a competitive observational
signal \cite{trispectrum1,trispectrum2}, we have
 focused only on the bispectrum which is characterised by the parameter
$f\sub{NL}$. In all cases our treatment
is based on existing ones, though we do not always agree with the original
authors.

The preheating and inhomogeneous reheating scenarios   cover
a  range of possibilities, which have not been fully explored but which 
can presumably allow a wide range for $f\sub{NL}$.
The same is true of multi-component inflation, except that 
extremely large values comparable with the current bound $|f\sub{NL}|\lsim
10^2$  seem relatively unlikely. In contrast, the simplest curvaton
scenario can produce a strongly negative value (even violating the 
current bound). However, in the important special case where 
the curvaton dominates the energy density before it decays, it
gives precisely $f\sub{NL}=+5/4$. Finally,
for the single-component inflation case,  Maldacena's calculation
combined with  current constraints on the spectral
tilt show that 
it has magnitude less than $10^{-2}$. These result are summarised in
the Table \ref{tscenarios}.
\begin{center}
\begin{table}[h]
\caption{Non-gaussianity according to different scenarios
for the creation of the curvature perturbation. For the simplest curvaton scenario,
 $f\sub{NL}= +5/4$ is a favoured value. \label{tscenarios}}
\begin{tabular}{lllll}
Scenario & $|f\sub{NL}|\ll 1$ & $|f\sub{NL}|\simeq 1$
&  $f\sub{NL}\ll -1$ &  $f\sub{NL}\gg 1$ \\
\hline
Single-component &&&& \\
inflation  & yes & no & no & no \\
Multi-component &&&& \\
inflation & likely & possible  & possible & possible  \\
Simplest &&&& \\
curvaton scenario & unlikely & likely
 & likely & no \\
\hline
\end{tabular}
\end{table}
\end{center}


In the near future, results from the Wilkinson Microwave Anisotropy Probe (WMAP) \cite{wmap} or elsewhere
 may detect a value $|f\sub{NL}|\gg1$.
If that does not happen, then PLANCK \cite{planck} or a successor will either detect
a value  $|f\sub{NL}|\sim 1$, or place a bound  $|f\sub{NL}|\lsim 1$.
The precise level at which this will be possible has yet to be determined
because it  would require a second-order calculation of all relevant
 observational signatures.  The example of the simplest curvaton scenario, 
where $f\sub{NL}
=+5/4$ is a favoured value, shows that such a calculation and the 
eventual observations will be well worthwhile.

\begin{acknowledgments}
We thank Karim A. Malik for discussions on all aspects of this work,
and Kari Enqvist, Asko Jokinen, Anupam Mazumdar, Tuomas Multam$\ddot{{\rm a}}$ki,
and  Antti V$\ddot{{\rm a}}$ihk$\ddot{{\rm o}}$nen
for a clarification of their work. 
We also
thank Misao Sasaki and Paolo Creminelli for discussions.
D.H.L. is  supported by PPARC grants PPA/G/O/2002/00469,  PPA/V/S/2003/00104,
PPA/G/S/2002/00098 and PPA/Y/S/2002/00272, and by European Union grant MRTN-CT-2004-503369. 
Y.R. is fully supported by COLCIENCIAS (COLOMBIA), 
and partially supported by COLFUTURO (COLOMBIA), UNIVERSITIES UK (UK), 
and the Department of Physics of Lancaster University (UK).
\end{acknowledgments}

\end{document}